\documentclass[amsmath,superscriptaddress,reprint,aps]{revtex4-2}
\usepackage{graphicx}
\usepackage{dcolumn}
\usepackage{bm}
\usepackage{hyperref}
\usepackage{siunitx}
\usepackage{color, soul}

\setstcolor{red}
\usepackage[mathlines]{lineno}

\begin{document}
	
	
	\title{Mechanically actuated Kerr soliton microcombs}

	\author{Shun~Fujii}
	\email[Corresponding author. ]{shun.fujii@phys.keio.ac.jp}
 	\affiliation{Department of Physics, Faculty of Science and Technology, Keio University, Yokohama, 223-8522, Japan}

	\author{Koshiro~Wada}
	\affiliation{Department of Electronics and Electrical Engineering, Faculty of Science and Technology, Keio University, Yokohama, 223-8522, Japan}
	
	
    \author{Soma~Kogure}
	\affiliation{Department of Electronics and Electrical Engineering, Faculty of Science and Technology, Keio University, Yokohama, 223-8522, Japan}

    \author{Takasumi~Tanabe}
	\affiliation{Department of Electronics and Electrical Engineering, Faculty of Science and Technology, Keio University, Yokohama, 223-8522, Japan}

	
\begin{abstract} 
Mode-locked ultrashort pulse sources with a repetition rate of up to several tens of gigahertz greatly facilitate versatile photonic applications such as frequency synthesis, metrology, radar, and optical communications. Dissipative Kerr soliton microcombs provide an attractive solution as a broadband, high-repetition-rate compact laser system in this context. However, its operation usually requires sophisticated pump laser control to initiate and stabilize the soliton microcombs, particularly in millimeter-sized ultrahigh-Q whispering-gallery resonators. Here, we realize a mechanically actuated soliton microcomb oscillator with a microwave repetition rate of 15 GHz. This enables direct soliton initiation, long-term stabilization, and fine tuning, where the operation now lifts the prerequisite pump laser tunability that must be relaxed if the technology is to be widely used outside the laboratory environment. We reveal the prospects for using this method with a wide range of applications that would benefit from mechanical soliton actuation such as optical clocks, spectral extension, and dual-comb spectroscopy.
\end{abstract}

	\maketitle
	

	\section*{Introduction}	
 
Optical frequency combs based on high-quality factor microresonators, known as `microcombs'~\cite{Kippenberg2018,Gaeta2019}, provide rich properties that are of significant importance in relation to immense frequency comb applications thanks to their compactness, high repetition rate microwave to terahertz operating range, and broad optical spectrum covering up to one octave. In particular, a dissipative Kerr soliton (DKS), which propagates while sustaining its waveform inside a nonlinear resonator, offers a luminous landscape for studying nonlinear optical phenomena as well as various intriguing applications~\cite{Sun2023}. In fact, DKSs existing in an anomalous dispersion system and recently discovered stable forms of microcombs (i.e., dark pulses~\cite{xue2015mode}, soliton crystals~\cite{Cole2017}) have been used for a number of proof-of-concept and more practical demonstrations. These include a phase-coherent link in the optical to microwave domain~\cite{Spencer2018}, high-capacity telecommunication~\cite{Marin-Palomo2017,Corcoran2020,Fueloep2018,Fujii:22}, photonic processing~\cite{Bai2023,Feldmann2021}, dual-comb ranging~\cite{Suh884,Trocha887} and spectroscopy~\cite{Suh600,Lucas2018}, and ultra-low noise microwave generation~\cite{Liang2015:high,Lucas2020,Yang2021,Kwon2022,Yao2022}. These applications are expected to fully exploit the intrinsic properties of microcombs that directly operate at tens to several hundreds of GHz, and have high-purity repetition rates with low-power continuous-wave (CW) pumping.

	\begin{figure*}
	\centering
		\includegraphics[width=\textwidth]{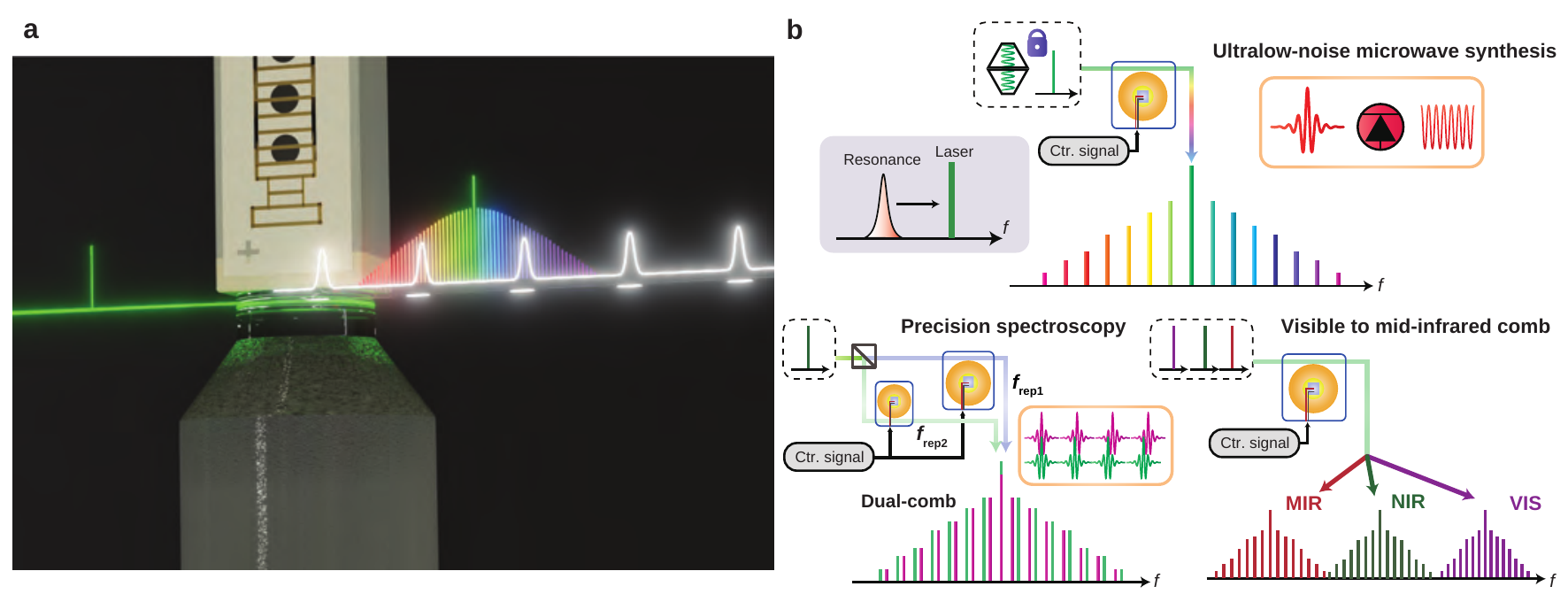}
		\caption{\label{fig1} Conceptual schematics and potential applications of mechanically actuated dissipative Kerr soliton (DKS) microcombs in a crystalline whispering-gallery (WG) mode resonator. (a) Mechanical actuation enables direct soliton generation without pump laser control. An ultrahigh-Q crystalline resonator emits a high-repetition rate pulse train with a continuous-wave (CW) input once an external modulation is applied to a piezo device attached to the resonator. (b) Examples of potential applications and deployment of mechanically actuated soliton microcomb sources. The frequency-referenced, ultra-stable CW laser can be directly used as a pump laser, and the mechanical channel provides an additional feedback control option for spectral purification or long-term stabilization after soliton initiation. The comb repetition frequency can also be tuned and modulated. In principle, the cavity actuation makes it possible to extend microcomb operation from the visible to the mid-infrared wavelength band because the set-up eliminates the need for high-performance wavelength-tunable lasers operating in such wavelength bands. Mechanical actuation is also expected to enhance the functionality of precision spectroscopy, where dual-combs with a slightly different repetition rates are generated from the same pump laser thanks to the independent actuation and resonance control of two resonators. This configuration naturally improves the mutual coherence between two pulses (green and purple) and the versatile tunability of a microcomb-based dual-comb system.}
	\end{figure*}


The ability to control oscillation frequencies and repetition rates are yet critical in microcomb systems. For example, the fine-tuning of comb frequencies and robust stabilization to external frequency references are central strategies for boosting the performance of optical-to-microwave frequency division~\cite{PhysRevLett.86.3288} and dual-comb applications~\cite{Coddington2016}. Despite the significant demand to pursue the tuning capability of soliton microcombs, the ultimately elemental configuration of optical microresonators made of single, dielectric materials poses a key challenge as regards achieving cavity actuation. The limited choice of available actuators not only reduces the tunability of microcomb frequencies but also increases system complexity in terms of operation. This issue must be addressed with regard to specific future applications and integration.

While several attempts have been made to establish a reliable way of tuning and stabilizing DKSs~\cite{Xue:16,Jung:14,PhysRevX.3.031003,Fujii2023}, the direct initiation of DKSs without pump laser control remains a vital challenge. Recent advances in the monolithic integration of piezoelectric actuators~\cite{Liu2020} and microheaters~\cite{Joshi:16} on silicon nitride (Si$_3$N$_4$) microring resonators have enabled the direct generation of $\sim$200~GHz mode spacing soliton microcombs and the high-speed, versatile control of comb properties. Carrier injection also allows the generation of coherent microcomb states in graphene-implemented silicon nitride resonators~\cite{Yao2018}, while photonic integrated resonators based on Pockels materials such as aluminum nitride (AlN)~\cite{Jung:14} and lithium niobate (LiNbO$_3$)~\cite{He2019,Wang2019} have revealed the potential to control microcomb states without pump laser control. In addition, the recent development of laser self-injection locking readily enables a turn-key operation for microcombs~\cite{Shen2020,Xiang2021} in fully-integrated photonic circuits with CMOS-compatible fabrication. Despite some demonstrations of the frequency stabilization of 'non'-phase-locked microcombs~\cite{PhysRevX.3.031003,7934411} or the frequency tuning and modulation of nonlinear optical processes~\cite{Werner2017}, direct soliton initiation and actuation remain key goals for monolithic ultrahigh-Q whispering gallery (WG) resonators.

This fact may limit the potential of ultrahigh-Q WG resonators as a leading platform for microcomb generation compared with integrated resonators. This is despite the considerable potential for many intriguing applications and demonstrations that exploit the unique properties of WG resonators, such as ultrahigh-Q across a broad transmission window~\cite{Chen2020,Lecaplain2016}, a diversity of available transverse modes~\cite{Lucas2018,PhysRevX.7.041055,Weng2020}, a variety of resonator shapes and materials~\cite{Kovach2020,Lin2019}, and easy fabrication and good functionality~\cite{Kovach2020,Tan2021}.
    
Here, we first demonstrate the direct initiation and frequency stabilization of soliton microcombs with a microwave repetition rate of 15 GHz in a mechanically-actuated WG crystalline microresonator. A commercial piezoelectrical lead-zirconate-titanate (PZT) element is used to actuate an ultrahigh-Q magnesium fluoride ($\mathrm{MgF_2}$) resonator via both mechanical deformation and elasto-optic effects, which allow a tuning efficiency of up to several tens of MHz/V and a kilohertz actuation bandwidth large enough for Pound-Drever-Hall (PDH) soliton stabilization. The dynamic modulation of the soliton repetition rate has also been demonstrated by utilizing a mechanical channel while maintaining pump laser detuning. These schemes eliminate the need for a high-speed wavelength-tunable laser in a complicated experimental setup as well as providing an additional feedback option that can be used for the fine tuning and stabilization of the comb frequencies. This simple and powerful working principle is universally applicable to other WG resonator platforms (e.g., silica and lithium niobate), and exploits the capability of optical parametric oscillators~\cite{Fujii:19,Sayson2019} and various microcomb applications, especially at microwave rates, such as ultra-low noise frequency synthesis~\cite{Spencer2018,Liang2015:high,Lucas2020,Yang2021,Kwon2022,Yao2022}, dual-comb spectroscopy, and optical clocks.

	\begin{figure*}
	\centering
		\includegraphics[width=\textwidth]{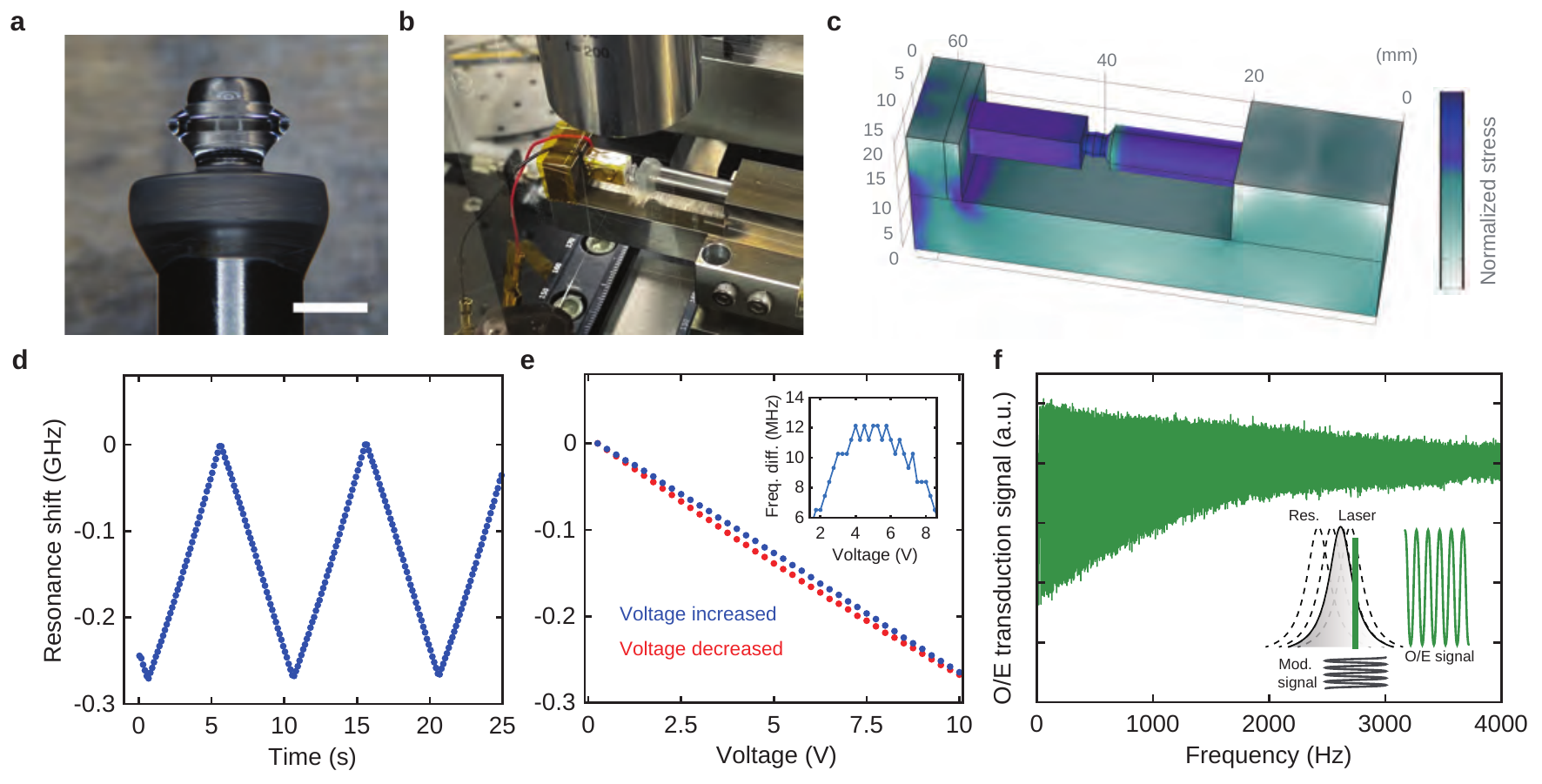}
		\caption{\label{fig2} Capacity for piezoelectric control of resonance frequencies. (a) Optical micrograph of a whispering-gallery (WG) mode magnesium fluoride crystalline microresonator. The scale bar represents 3 mm. (b) Photograph of the optical measurement setup and a mounted resonator, where a piezoelectric transducer (PZT) element induces mechanical stress on the flat-top region of the resonator. (c) Real-scale simulation of mechanical stress in our device. A large stress is observed in the $\mathrm{MgF_2}$ resonator region fixed between a PZT element and a stainless steel rod, enabling wide mechanical frequency tuning. (d,e) Observed resonance frequency shift with a slow PZT scan with a 0.1~Hz ramp function whose peak-to-peak voltage is 10~V and observed weak hysteresis behavior of mechanical resonance tuning. The blue dots (red dots) correspond to the frequency shift when increasing (decreasing) the applied voltage. The inset shows the frequency difference between the upward and downward directions, yielding a maximum difference of approximately 12~MHz at 5~V. (f) Measured frequency response of a certain resonance. An optical-to-electrical (O/E) transduction signal is obtained by recording the cavity transmittance while varying the modulation frequency applied to the PZT element. The inset shows the working principle of the measurement.}
	\end{figure*}
	
\section*{Results}
\subsection*{Concept and device characterization}
Schematics of the working principle and potential deployment of mechanically-actuated DKS generation are shown in Fig.~\ref{fig1}. A piezoelectric (PZT) element attached on top of the resonator is used to trigger DKS generation via cavity resonance scanning. The resonance modes are actuated by both mechanical deformation and the elasto-optic effect when an electrical signal is applied to the electrodes of a PZT device. Because the resonant frequency can be tuned by $\sim$1000 times the cavity linewidth ($\sim$100~kHz) with only a few volts, it is easy to scan the resonant frequency from the blue- to the red-detuned region (i.e., the soliton regime) across a fixed pump laser frequency in the same way as with conventional pump laser scanning~\cite{Herr2014}. With a continuous-wave (CW) pumping above a threshold, a mode-locked GHz-range repetition rate pulse is generated via Kerr-nonlinearity and then detected in the optical and electrical domains. The resonance control via a mechanical channel offers various potential applications and improves the capability of high-repetition rate microcombs as discussed later.

We first investigate the capability of the piezoelectric resonance tuning of our devices. A crystalline microresonator with a free-spectral range (FSR) of 15.24~GHz and a Q-factor of $10^9$ is fabricated by manual shaping and surface polishing such that the resonator has a flat top region that contacts uniformly with a PZT device (Fig.~\ref{fig2}(a)). The resonator and PZT actuator are mounted in a carefully designed U-shaped jig to apply a proper load to the PZT actuator and mechanical stress to the resonator, as shown Fig.~\ref{fig2}(b) and \ref{fig2}(c). The tuning efficiency of the resonant frequencies and the mode-spacing, i.e., the FSR, is expressed as, $df/dV = -(dF/dV) (\nu/\pi E R^2) f$ and $df_\mathrm{FSR}/dV= -(dF/dV) (\nu/\pi E R^2) f_\mathrm{FSR}$, respectively, where $dF/dV$ represents the force per applied voltage, $R$ is the effective radius of the resonator, $\nu$ is Poisson’s ratio, and $E$ is Young’s modulus. It should be noted that the above relation neglects the elasto-optic effect because the mechanical deformation, i.e. pure geometrical change in the resonator shape, is considered to constitute a dominant contribution in the resonant mode scan in this study. (A theoretical description and a detailed discussion of tuning efficiency are presented in Supplementary Note 1).

 	\begin{figure*}
	\centering
		\includegraphics[width=\textwidth]{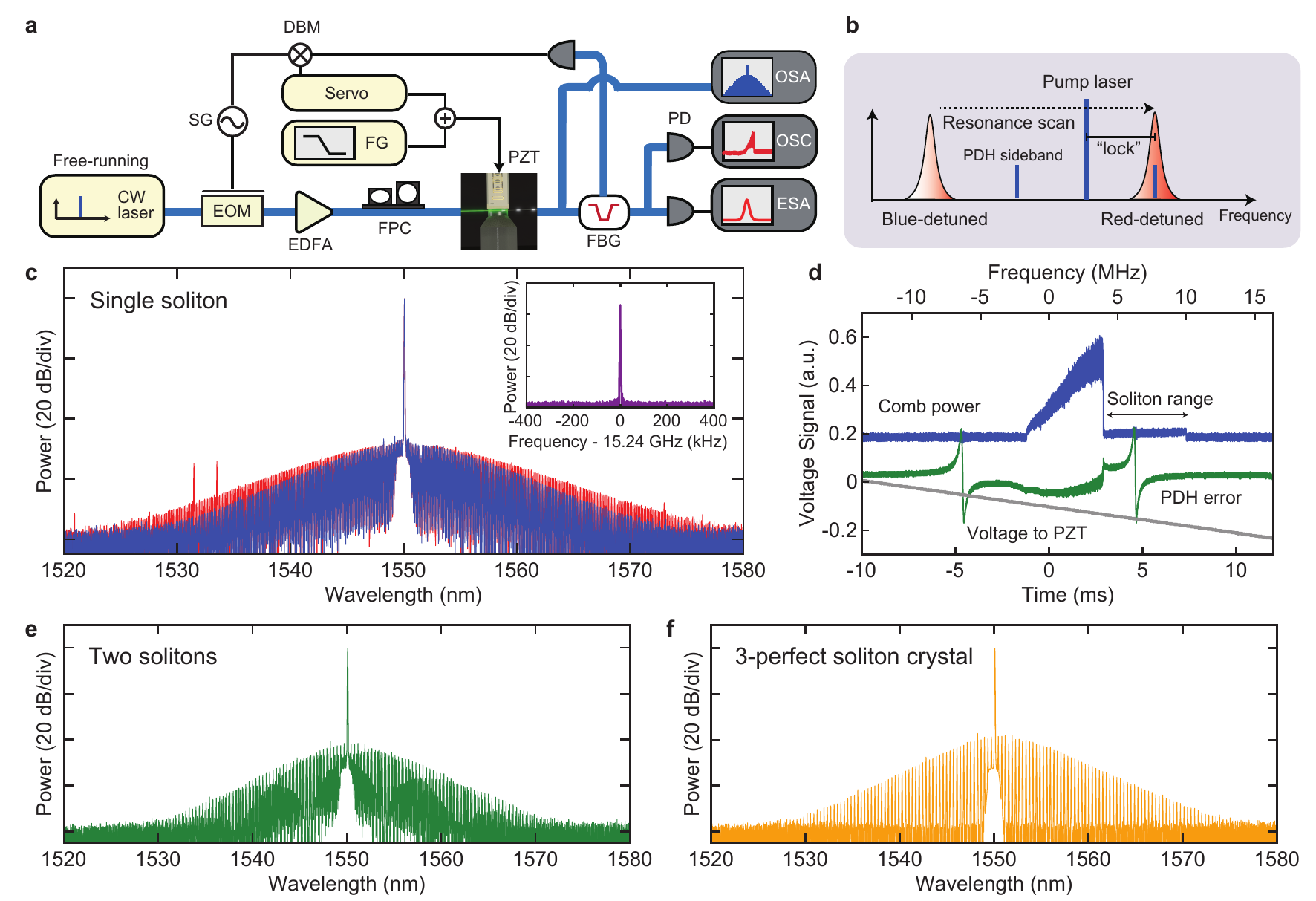}
		\caption{\label{fig3} Mechanically initiated soliton microcombs. (a, b) Experimental setup and schematics of mechanically actuated soliton generation and stabilization. EOM, electro-optic modulator; SG, signal generator; EDFA, erbium-doped fiber amplifier; FPC, fiber polarization controller; FBG, fiber Bragg grating filter; PD, photodetector; DBM, double-balanced mixer; ESA, electrical spectrum analyzer; OSC, oscilloscope; OSA, optical spectrum analyzer; FG, function generator; PZT, piezoelectric transducer. The resonance frequency is scanned across a CW laser instead of using conventional laser scanning. (c) Optical spectra of a single-soliton state with different detuning at $\delta \omega/2\pi =$ 5~MHz (blue) and 7.5~MHz (red). The inset shows a radio-frequency beat-note spectrum supporting soliton mode-locking at a repetition rate of $\sim$15.24~GHz. (d) Typical monitored comb power (blue) associated with a soliton step while scanning the voltage applied to a PZT device (gray). Decreasing voltage induces resonance scanning in the high-frequency direction. The high-frequency sideband of a PDH error signal (green) is used for soliton stabilization. The locking point is controlled by changing the frequency and phase of a modulation signal ($\delta \omega/2\pi =$ 6.3~MHz in this case). (e,f) Measured optical spectra of a two-soliton state and three-perfect soliton crystal (PSC) state. }
	\end{figure*}

The tuning efficiency is measured by optical heterodyne detection with a reference laser. After locking the pump laser to a resonance mode via the PDH method, the resonance frequency is slowly scanned (0.1~Hz) by applying a ramp voltage ($V_\mathrm{pp}=10$~V) to the PZT device. This enables the bidirectional measurement of resonance shift versus applied voltage as displayed in Fig.~\ref{fig2}(d). The resonance shift exhibits a weak non-linear hysteresis behavior inherent to the PZT device (Fig.~\ref{fig2}(e)); the inset shows the frequency difference between upward and downward directions, yielding a maximum difference of 12~MHz at 5~V, where the frequency resolution is limited by the bandwidth setting of the electrical spectrum analyzer. The extracted linear tuning efficiency is $df/dV = -27$~MHz/V although the predicted theoretical value is $\sim-314$~MHz/V, which yields an efficiency of $\sim$8.6~\% (calculated by dividing the experimental value by the theoretical prediction). Experimentally, the typical efficiency ranges from 3~\% to 15~\%, and the discrepancy is mainly due to a low mechanical connection. In fact, the loading condition and stiffness of a U-shaped jig substantially affect the tuning efficiency and bandwidth as discussed later. Even though the tuning efficiency can also be evaluated with a transmission measurement, the above method features fast and highly accurate measurement, where the frequency resolution is determined by either the spectrum analyzer or the frequency stability of the reference laser  (See Methods and Supplementary Note 2 for details and extended results).

Figure~\ref{fig2}(f) shows the result of the frequency response measurement. The optical-to-electrical (O/E) transduction signal is obtained by recording the cavity transmittance while varying the modulation frequency, which directly yields the O/E converted frequency response as indicated in the inset of Fig.~\ref{fig2}(f). The peak-to-peak voltage of the transmittance signal gradually decays as the modulation frequency increases, and this results in a 3 dB-bandwidth of approximately 1.5~kHz. These properties guarantee a resonant scan speed of up to $\sim$100~MHz/ms when applying a 0-5 V ramp voltage, which is fast enough for visualizing a clear soliton step in ultrahigh-Q resonators. The experimental setups are detailed in Methods and Supplementary Note 3.

 	\begin{figure*}
	\centering
		\includegraphics[width=\textwidth]{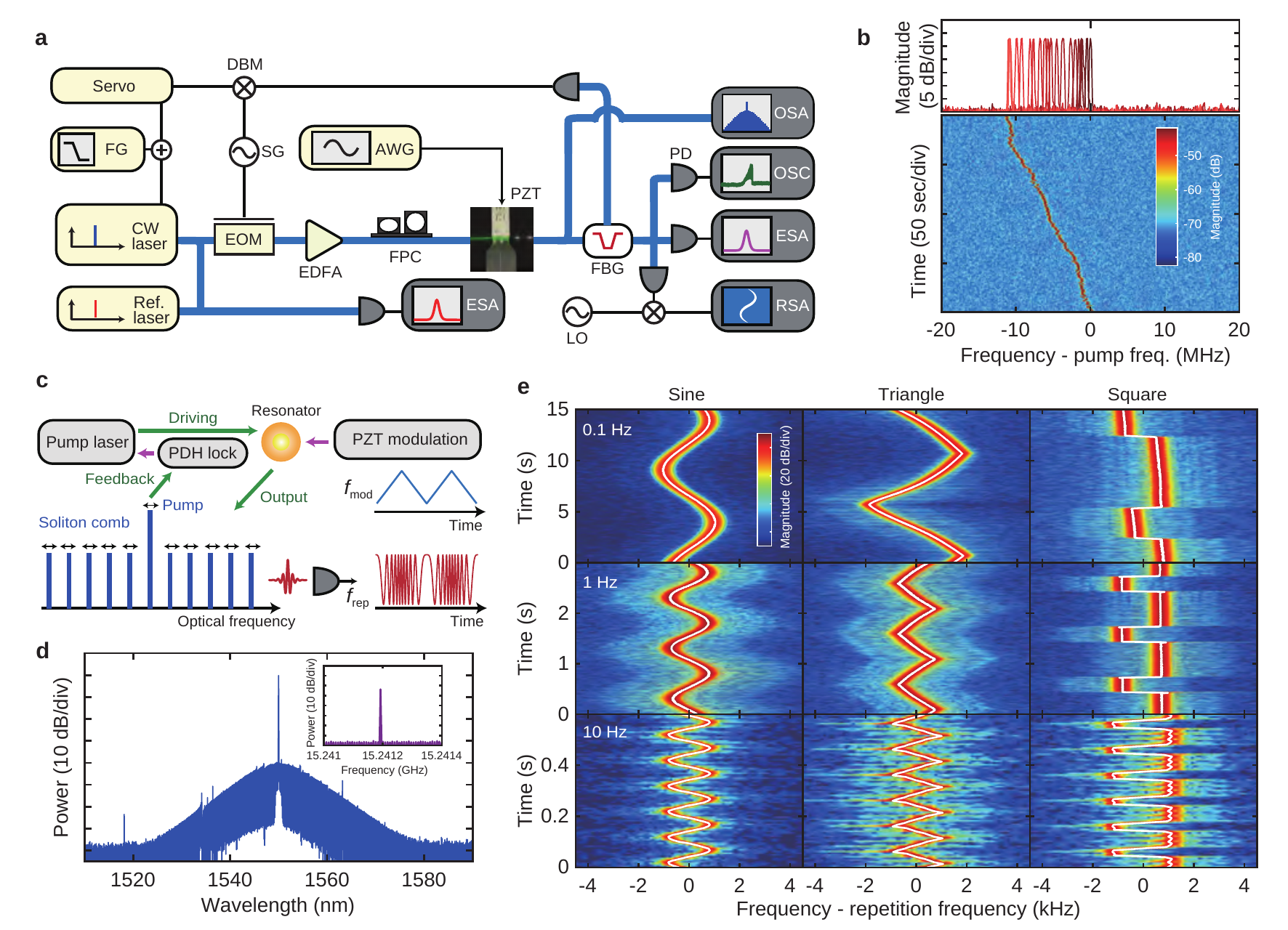}
		\caption{\label{fig4} Dynamic arbitrary repetition-rate modulation of detuning-stabilized soliton microcombs. (a) Experimental setup for soliton repetition-rate modulation and frequency drift measurement. Ref. laser: Reference laser used for beat-note measurements; RSA, real-time spectrum analyzer; LO, local oscillator; AWG, arbitrary waveform generator. The feedback signal via the PDH circuit is sent to the pump laser to maintain the pump detuning during the repetition rate modulation. (b) Soliton (center) frequency drift recorded by a beat-note measurement of the pump laser and reference laser after soliton formation. The frequency gradually drifts by up to 10~MHz in 200 seconds due to the temperature fluctuation of the pump mode frequency. (c) Conceptual diagram of mechanically actuated repetition rate modulation combined with pump detuning stabilization via PDH locking. The soliton repetition frequency is monitored by using an RSA via frequency down-mixing with a local oscillator. (d) Optical spectrum of a single soliton state with a repetition rate of 15.24 GHz when the pump detuning is stabilized at 10 MHz. The inset shows the beat-note spectrum. (e) Real-time characterization of arbitrary soliton repetition rate modulation. The modulation waveforms (sinusoidal, triangular, square) and frequencies (0.1, 1, 10~Hz) can be controlled with an AWG. The amplitude of the modulation voltages and the duty cycle of the square waveforms are set at 0-2 V and 0.3, respectively.}
	\end{figure*}

 \subsection*{Mechanically-actuated soliton initiation and stabilization}

As the tuning bandwidth is large enough to sweep the resonance frequency, we next demonstrate soliton initiation and stabilization via piezoelectric mechanical control. The experimental setup and schematics are shown in Fig.~\ref{fig3}(a) and \ref{fig3}(b). First, we search for a soliton mode by electrically driving a PZT device using a function generator (FG) (i.e., sweeping resonance frequency) instead of laser frequency scanning. When the total tuning range is too short to find the desired soliton mode (typically up to a few GHz at maximum even when using a high-voltage amplifier), wide-range thermal tuning~\cite{Fujii2023} can be implemented for a wide-range search exceeding the cavity FSR. We emphasize again that no pump laser control is needed in this experiment.

For soliton generation, a free-running CW laser source is amplified by an erbium-doped fiber amplifier, and an optical power of $\sim$200~mW is coupled via a tapered fiber. Reducing the applied voltage induces a resonance shift in the higher frequency (shorter wavelength) direction (Fig.~\ref{fig3}(b)). Figure~\ref{fig3}(c) shows observed single-soliton states, where the pump detuning (=$\delta \omega/2\pi$) is slightly changed by varying the PZT voltage. The inset indicates the soliton repetition rate of 15.24~GHz. To confirm that the PZT voltage is effective in controlling the comb bandwidth and long-term detuning stabilization, a servo output is then fed back to the PZT device to compensate for the long-term frequency drift of both the resonance and the pump laser that substantially causes solitons to disappear. A typical PDH error signal and soliton steps are shown in Fig.~\ref{fig3}(d); a high-frequency sideband of a PDH error signal is used for the detuning stabilization~\cite{PhysRevLett.121.063902,Fujii2023}. Once the feedback loop is engaged, the comb bandwidth is successfully fixed thanks to its fast mechanical frequency response. Multi-solitons are stochastically observed as shown in Fig.~\ref{fig3}(e) and \ref{fig3}(f), corresponding to the 2-soliton state and 3-perfect soliton crystal (PSC), respectively. The multi-soliton states can be deterministically settled in a single soliton state~\cite{Guo2017} via backward resonance tuning by changing the PZT voltage.

 \subsection*{Dynamics of arbitrary soliton repetition-rate modulation}
 We next demonstrate arbitrary repetition rate tuning via the mechanical response. Since the soliton repetition rate varies even when the effective detuning changes, the detuning scan inevitably introduces the simultaneous variation of the soliton power and bandwidth. A feed-forward scheme~\cite{Liu2020}, which modulates both the pump laser frequency and resonance frequency, is suitable for several applications, whereas the nature of the ultrahigh-Q resonance ($>10^9$) in crystalline resonators typically narrows the soliton existence range to less than $\sim$10~MHz, and this leads easily to the abrupt decay of soliton states unless we employ detuning stabilization~\cite{Yi:16,Lucas2017}. Therefore, we perform repetition rate modulation while maintaining detuning stabilization as shown in Fig.~\ref{fig4}(a). The PDH feedback loop is applied to the frequency modulation of the pump laser for detuning stabilization, and an arbitrary waveform generator (AWG) is used to modulate the soliton repetition rate. In this scheme, the deviation of the pump frequency is monitored by observing the beat signal of the pump laser and the reference CW laser as shown in Fig.~\ref{fig4}(b), which indicates that the frequency has drifted up to 10~MHz in 200 seconds. The frequency drift is mainly attributed to the thermal fluctuation of the microresonator resonance in which a 1~mK temperature fluctuation causes a $\sim$1.8~MHz frequency shift in $\mathrm{MgF_2}$ crystalline resonators~\cite{Fujii2023}. The thermal stability can be improved by operating the system with sub-mK temperature control and compact packaging~\cite{Lim2019}.
 
The working principle of mechanical repetition-rate modulation with detuning stabilization is shown in Fig.~\ref{fig4}(c). The detected microwave repetition rate of a single soliton comb (Fig.~\ref{fig4}(d)) is mixed with the local oscillator (LO) frequency, and the down-mixed signal is monitored by using a real-time spectrum analyzer (RSA). We demonstrate the arbitrary repetition rate modulation as a proof of concept with general multi-functions (i.e., sinusoidal, triangular, and square) at different modulation periods. Figure~\ref{fig4}(e) indicates that the dynamic change in repetition rates is approximately 2.5-4.0~kHz with a peak-to-peak voltage of 2~V, which is consistent with the predicted value of 3.9~kHz under the assumption of a mechanical connection efficiency of 8~\%. As seen in the spectrograms, the modulated repetition frequencies exhibit weak dependence on the function type and period. This is likely because of the unexpected nonlinear behavior of the mechanical responses, but further investigation is required. The gradual deviation of repetition rates from the offset frequency is due to the long-term temperature fluctuation and is not caused by the frequency modulation. Although the modulation frequency is limited by the refresh rate and frequency resolution of our RSA, we confirmed that the modulation frequency can be increased up to a few kHz without losing soliton states. It should be noted that the frequency response also depends on the locking bandwidth of the detuning stabilization fed back to the pump laser in this experiment.

\section*{Discussion}

The presented results mark the first demonstration of mechanically-actuated soliton initiation and stabilization, and the dynamic modulation of soliton repetition rates in ultrahigh-Q WG microresonators. Our approach inspires several important advances in soliton microcomb applications and technologies. First, this scheme in principle enables simple, cost-effective operation with a compact package comprising soliton microcomb modules with a gigahertz repetition rate. This is because a robust system consisting of a PZT element and a U-shaped metal jig replaces expensive, bulky, frequency-tunable lasers that are used to access a soliton state with conventional laser sweeping. Second, soliton generation is readily possible with a frequency-referenced CW laser or a sub-Hz-linewidth ultra-stable laser as a pump source, which can significantly reduce microcomb noise and results in ultrahigh-purity photonic microwave generation~\cite{Lucas2020}. Mechanical actuation also provides additional control of frequency comb properties such as repetition frequency and carrier-offset envelope frequency. This makes it possible to realize a more advanced phase-coherent optical link with frequency standards by combining with other control techniques~\cite{Kwon2022,Lucas2020,PhysRevLett.121.063902} (e.g., pump frequency, pump power). Moreover, this method removes the limitation imposed by the available wavelength bands of narrow-linewidth tunable lasers and optical components, which has been hindering microcomb generation (not limited to soliton states) in important but challenging wavelength bands (e.g., visible, 1.06 \textmu m, and mid-infrared). With proper resonator dispersion engineering~\cite{FujiiTanabe+2020+1087+1104,Fujii:20} and spectral extension techniques~\cite{Jost:15,Zhang2020,Moille2021}, a self-referencing scheme could be implemented in the future. Furthermore, dynamic repetition rate modulation offers useful features for some potential applications such as dual-comb spectroscopy and optical sampling.

Since the actuation induces a slight change in the radial size of the resonator, the system is highly compatible with an all free-space system with a prism coupler as a way of controlling the coupling rate, and this could be combined with attractive laser self-injection locking techniques~\cite{Pavlov2018}. Technically, it is possible to integrate a piezo actuator in ring-like WG microresonators~\cite{Werner2017}, which makes the system simpler than the current configuration. Here, we used $\mathrm{MgF_2}$ crystalline microresonators as a platform, but other WG resonators such as silica rods and lithium niobate disks can be employed in the same manner. As well as the mechanical properties of the materials (i.e., Poisson's ratio, Young's modulus), the material and geometry of a U-shaped jig is also a non-trivial factor that determines the mechanical connection efficiency. Generally, it is reasonable to choose material for a jig that is stiffer than the resonator material. We compared the tuning efficiency with two different jigs made of stainless steel ($E = 205$~GPa) and aluminum ($E = 70$~GPa) (cf. $E = 139$~GPa for $\mathrm{MgF_2}$ crystal), and the stainless-steel jig had a higher tuning efficiency than the aluminum one. The results are detailed in Supplementary Note 4. As a principal guideline, we need to employ a piezo actuator with a large $dF/dV$, a smaller resonator ($1/\pi R^2$), and soft resonator material ($\nu/E$) to increase the frequency tuning efficiency. Moreover, we expect that further improvement of the jig design or direct-bonding structures will refine the frequency response and tuning efficiency. The use of a finite-element simulation to analyze the mechanical eigenfrequency and applied stress also helps to further improve the design (See Methods and Supplementary Note 4 for detailed analyses). 

In summary, we have demonstrated mechanically-actuated direct initiation and precise control of X- and K-band (8-27~GHz) repetition-rate soliton microcombs in ultrahigh-Q crystalline microresonators. The demonstrated system adds excellent new features to ultrahigh-Q WG microresonator-based microcomb generation e.g. it is tunable laser-free and has high-bandwidth frequency control and stabilization that were previously lacking. We believe that this approach has promising potential for an ultralow-noise photonics microwave generator, dual-comb and distance measurement technologies, and for extending microcomb wavelengths in a simple, compact, and low-cost package.

	\section*{Acknowledgments}
This work is supported by JSPS KAKENHI (JP19H00873, JP22K14625) and Strategic Information and Communications R\&D Promotion Programme (SCOPE) (JP225003008) from the Ministry of Internal Affairs and Communications. S. F. thanks the support from Mizuho Foundation for the Promotion of Sciences. We are grateful for the technical support provided by the Manufacturing Center at Keio University. We sincerely thank H. Kumazaki for his technical assistance and advice, and K. Yube for supporting graphic sketch making.

	\section*{Author contributions}
S. F. designed and supervised the project. S.F and K.W. performed the experimental measurement. S.F. conducted the finite-element simulation. K. W. fabricated the crystalline microresonators, and S. F., K. W., and S. K. developed the experimental setups.  S. F wrote the manuscript with input from T. T. and all the authors contributed to constructive discussions.
	\bibliography{soliton_tuning}
\end{document}